\documentclass[letterpaper]{article}
\usepackage{aaai20}  % DO NOT CHANGE THIS
\usepackage{times}  % DO NOT CHANGE THIS
\usepackage{helvet} % DO NOT CHANGE THIS
\usepackage{courier}  % DO NOT CHANGE THIS
\usepackage[hyphens]{url}  % DO NOT CHANGE THIS
\usepackage{graphicx} % DO NOT CHANGE THIS
\usepackage{graphicx}  % DO NOT CHANGE THIS
\usepackage{amsmath}
\usepackage[ruled,linesnumbered]{algorithm2e}
\usepackage{subcaption}
\usepackage{multirow}
\usepackage{xcolor}

\title{Multiplex Graph Association Rules for Link Prediction}

\author{Michele Coscia\textsuperscript{\rm 1}, Michael Szell\textsuperscript{\rm 1,2}\\
\textsuperscript{\rm 1}IT University of Copenhagen\\
Rued Langgaards Vej 7\\
Copenhagen, Denmark 2300\\
mcos@itu.dk, misz@itu.dk\\
\textsuperscript{\rm 2}ISI Foundation\\
Via Chisola 5\\
Turin, Italy 10126
}

\begin{document}

\maketitle

\begin{abstract}
Multiplex networks allow us to study a variety of complex systems where nodes connect to each other in multiple ways, for example friend, family, and co-worker relations in social networks. Link prediction is the branch of network analysis allowing us to forecast the future status of a network: which new connections are the most likely to appear in the future? In multiplex link prediction we also ask: of which type? Because this last question is unanswerable with classical link prediction, here we investigate the use of graph association rules to inform multiplex link prediction. We derive such rules by identifying all frequent patterns in a network via multiplex graph mining, and then score each unobserved link's likelihood by finding the occurrences of each rule in the original network. Association rules add new abilities to multiplex link prediction: to predict new node arrivals, to consider higher order structures with four or more nodes, and to be memory efficient. In our experiments, we show that, exploiting graph association rules, we are able to achieve a prediction performance close to an ideal ensemble classifier. Further, we perform a case study on a signed multiplex network, showing how graph association rules can provide valuable insights to extend social balance theory.
\end{abstract}

\section{Introduction}
Complex networks are a powerful abstraction of interacting entities (nodes and links), well suited to study complex systems, from society, the brain, to interdependent infrastructure services. Given its analytical power, network analysis can be used to forecast the future status of a system, for instance to predict new relationships or routes. This is the well-known problem of link prediction: the task of estimating the likelihood of unobserved connections to be observed in the future \cite{liben2007link,lu2011link}. A powerful technique to solve link prediction in simple networks is graph association rules as in \texttt{GERM} \cite{berlingerio2009mining,bringmann2010learning}.

\begin{figure}
\centering
\includegraphics[width=.33\textwidth]{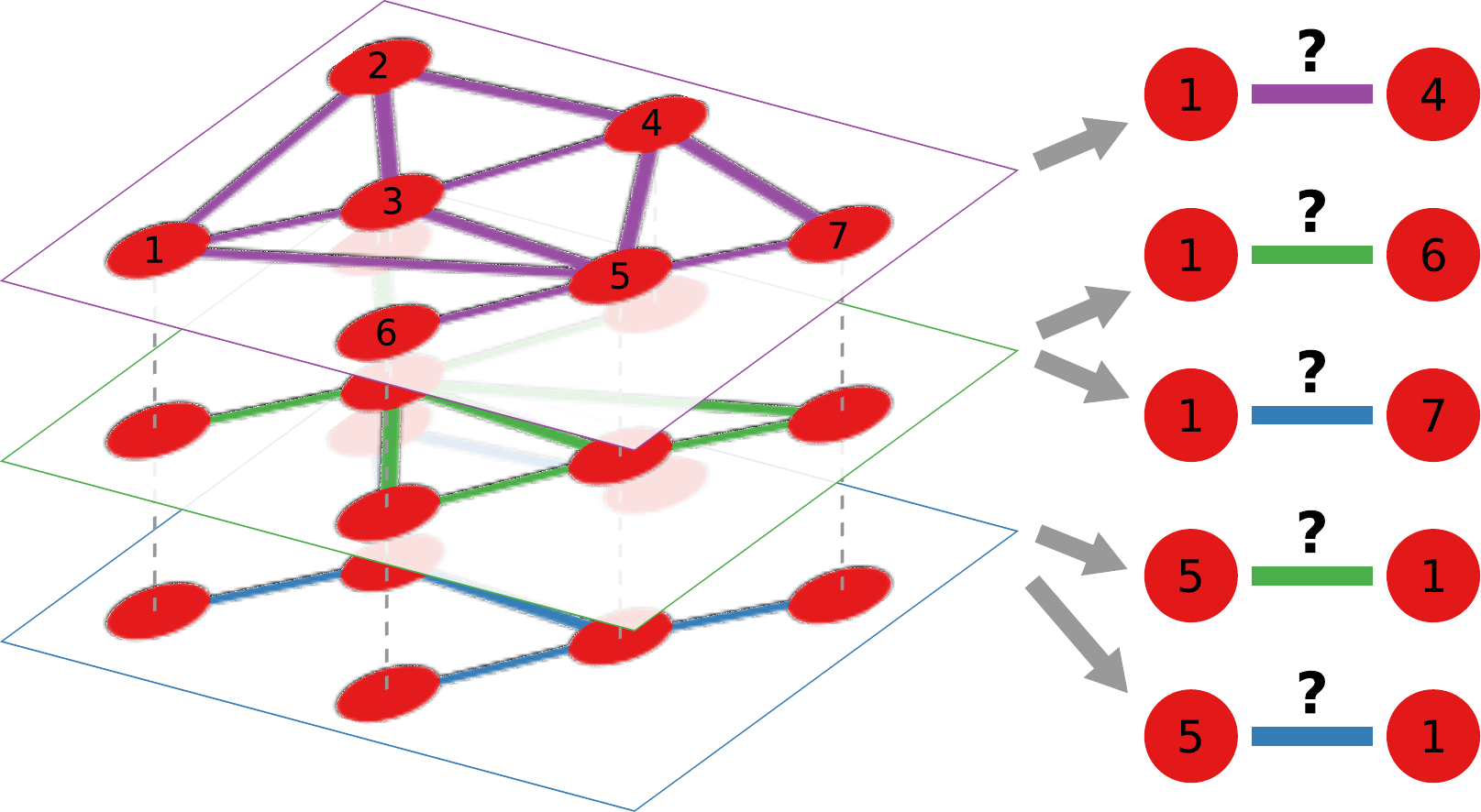}
\caption{A depiction of the multiplex link prediction problem. Will two nodes connect, and of which type?}
\label{fig:mllp}
\end{figure}

In multiplex networks, entities can connect in different ways \cite{krackhardt1987cognitive,roethlisberger1939management,kivela2014multilayer,boccaletti2014structure}. For example, you know people for different reasons -- friendship, work ties, economic transactions --, or you move through space with different means of transportation -- bicycle, car, train, plane \cite{berlingerio2011foundations,dickison2016multilayer}. Multiplex link prediction \cite{rossetti2011scalable,matsuno2018mell} comes with additional challenges, as Figure \ref{fig:mllp} shows. Here, we are not only interested in knowing \textit{whether} two nodes will connect: we also want to know \textit{how} they will connect. In fact, as the figure shows, nodes that are already connected are still part of the solution space: for example, nodes $1$ and $5$ are connected by a link from the top layer, but not from the middle and bottom ones. Thus, we want to estimate the likelihood they are going to connect via edges of multiple types -- or from multiple layers, since multiplex networks are a special subtype of multilayer networks (throughout the paper we use the terms ``type'' and ``layer'' interchangeably).

There are other approaches to perform multiplex link prediction \cite{pujari2015link,jalili2017link,sharma2016efficient,hristova2016multilayer,de2017community}. Most of them share a strategy: calculate classical link prediction scores based on the topology of the network, and then combine them into a multiplex score.

In this paper we extend the usage of graph association rules to perform link prediction on multiplex networks and prove their advantages over the alternatives. We need to significantly change \texttt{GERM}'s framework because graph association rules are based on frequent graph pattern mining, which can handle only a single label on the edges. \texttt{GERM} uses the attribute to indicate the link's appearance time, while here we need to use it to indicate its type.

First, we use \texttt{Moss} \cite{borgelt2005moss} to perform multiplex graph mining, i.e.~the discovery of all frequent multiplex patterns. Then, we use these patterns to build multiplex graph association rules, connecting two frequent patterns that differ by one link. Finally, we find all occurrences of a rule in the original graph and score the likelihood of the new link to appear. 

This approach has several advantages over the current state of the art in multiplex link prediction, which are the main contributions of the paper:

{\bf Higher order structures.} We are not limited to pairwise or three-way interactions as in classical link prediction. We can consider structures of four or more nodes. This is particularly relevant for social balance, where some edge types are considered positive and others are considered negative. Following the adage ``an enemy of my enemy is my friend'', some triadic closures are considered balanced and other unbalanced. In real world signed networks, balanced triangles are overexpressed and thus should be prioritized when performing link prediction. In this paper we show how triadic closure is overly simplistic when expanding from triangles to patterns of four nodes.

{\bf Links to new nodes.} We can predict not only where new links will appear, but also to where new nodes will connect, a feature not present in other multiplex link predictors.

{\bf Memory efficiency.} By not assigning a likelihood to every pair of unconnected nodes, we can be more memory efficient than  alternative predictors.

In our experiments, we show that, using graph association rules, we can achieve higher Area Under the ROC Curve (AUC) performance over several datasets representing systems coming from different fields, from web-mediated online social interactions to neural networks.

Note that we can use our framework to perform single layer link prediction via graph association rules. Our framework is open-source and freely available together with the data and code necessary to replicate our experiments\footnote{\url{http://www.michelecoscia.com/?page_id=1857}. The library also includes our implementations of other multiplex link prediction techniques. Note that, to properly run the code, you also need the \texttt{Moss} software, which we do not repackage. You can obtain it from \url{http://www.borgelt.net/moss.html}. We use the 6.15 (2016.07.05) version. Some baseline methods require external binaries, recoverable from \url{https://www.mapequation.org/} and \url{https://github.com/cdebacco/MultiTensor}.}.

\section{Related Work}
In this paper we address multiplex link prediction via mining graph association rules, implying that we need to perform frequent pattern mining over a multiplex network.

\subsubsection{Frequent Graph Pattern Mining}
Frequent pattern mining in graphs is the search for frequent subgraph patterns \cite{chakrabarti2006graph}. Originally, it was developed to find frequent patterns in a graph database that contains many small graphs. In this setting, the frequency (or support) is the number of graphs in the database containing the pattern. Among the most important algorithms are gSpan \cite{yan2002gspan,yan2003closegraph}, Gaston \cite{nijssen2004quickstart}, \texttt{Moss} \cite{borgelt2007canonical}.

In single graph mining, support is redefined as the number of times a pattern appears in a single graph. Naively counting the occurrences of a pattern breaks the anti-monotonicity requirement of the support \cite{kuramochi2005finding}: a larger pattern must have a support equal to or lower than the patterns it contains. If it does not, the search space cannot be efficiently pruned. For this reason, different definitions of support have been proposed \cite{kuramochi2005finding,fiedler2007support,bringmann2008frequent,elseidy2014grami,abdelhamid2016scalemine}.

Our paper extends the link prediction literature by exploiting \texttt{Moss}' \cite{borgelt2005moss} ability to perform pattern mining on multiplex networks, networks where nodes can be connected by multiple qualitatively different links \cite{berlingerio2011foundations,kivela2014multilayer,boccaletti2014structure,dickison2016multilayer}. Multiplex networks have been widely adopted in a variety of network analysis applications such as community discovery \cite{mucha2010community,berlingerio2011finding}, node ranking \cite{de2015ranking}, spreading processes \cite{de2016physics}, and even probabilistic motif analysis \cite{battiston2017multilayer}.

To the best of our knowledge, there are only three approaches that come close to multiplex graph pattern mining, each with its own downside: (i) a special case with only two layers \cite{bachi2012classifying} (signed networks), (ii) FANMOD \cite{wernicke2006fanmod}, included in Muxviz \cite{de2015muxviz}, which uses a non-monotonic support definition, and (iii) a subgraph mining approach \cite{anchuri2018mining}, which requires to provide the input patterns of interest -- and also has a non-monotonic support definition. None of these limitations apply to our proposed approach.

\subsubsection{Link Prediction}
In link prediction we observe a network at different moments in its evolution. The task is to estimate the likelihood of appearance of unobserved links \cite{liben2007link,lu2011link}. Most link predictors determine the link likelihood either using topological properties of the network -- thus they are unable to predict old-new links --, and/or operate on networks where nodes can connect to each other only via the same type of relation. Our approach has neither limitation.

First, we do not use topological measures, but we extract network motifs and we use them to build graph association rules. We base this part of our methodology on \texttt{GERM} \cite{berlingerio2009mining,bringmann2010learning}, and we improve over it by considering multiplex networks.

Second, we tackle multiplex link prediction, to predict the link type connecting two nodes \cite{rossetti2011scalable}. Although there are many approaches to this problem \cite{pujari2015link,jalili2017link,sharma2016efficient,hristova2016multilayer,de2017community}, they share the general idea of combining single layer scoring functions to consider inter-layer correlations.

Link prediction can be done via graph embeddings techniques \cite{goyal2018graph}. We know of one multilayer graph embedding technique \cite{li2018multi} which has not been used for multiplex link prediction, and another \cite{matsuno2018mell} which has. We leave this comparison for future work, noting that our approach is significantly different as it produces interpretable rules, rather than being a deep learning approach which exclusively focuses on maximizing the predictive performance.

\subsubsection{Applications}
Link prediction in general, and multiplex link prediction specifically, has a number of applications in many fields. Here we briefly discuss its relevance to the online social media community. One key aspect of online social media is the multiple identities of the same people across different platforms. To truly understand the spreading of information and social behaviors online, one needs to align social networks across platforms \cite{zhang2015multiple}: to identify the same users having profiles on Facebook, Twitter, etc. Some platforms might be harder to crawl than others, thus one could use multiplex link prediction to complete the information in one layer by extracting relevant rules from the other layers.

Other applications of multiplex link prediction for online social media include the analysis of brokerage between individuals in virtual and in-presence social networks \cite{hristova2015multilayer}; inform researchers on the privacy risks associated with the complex structural information embedded in multilayer networks \cite{rossi2015k}; and the planning of cross-platform marketing campaigns \cite{vikatos2020marketing}.

\section{Problem Definition}\label{sec:prob}

\begin{figure}
\centering
\includegraphics[width=.75\columnwidth]{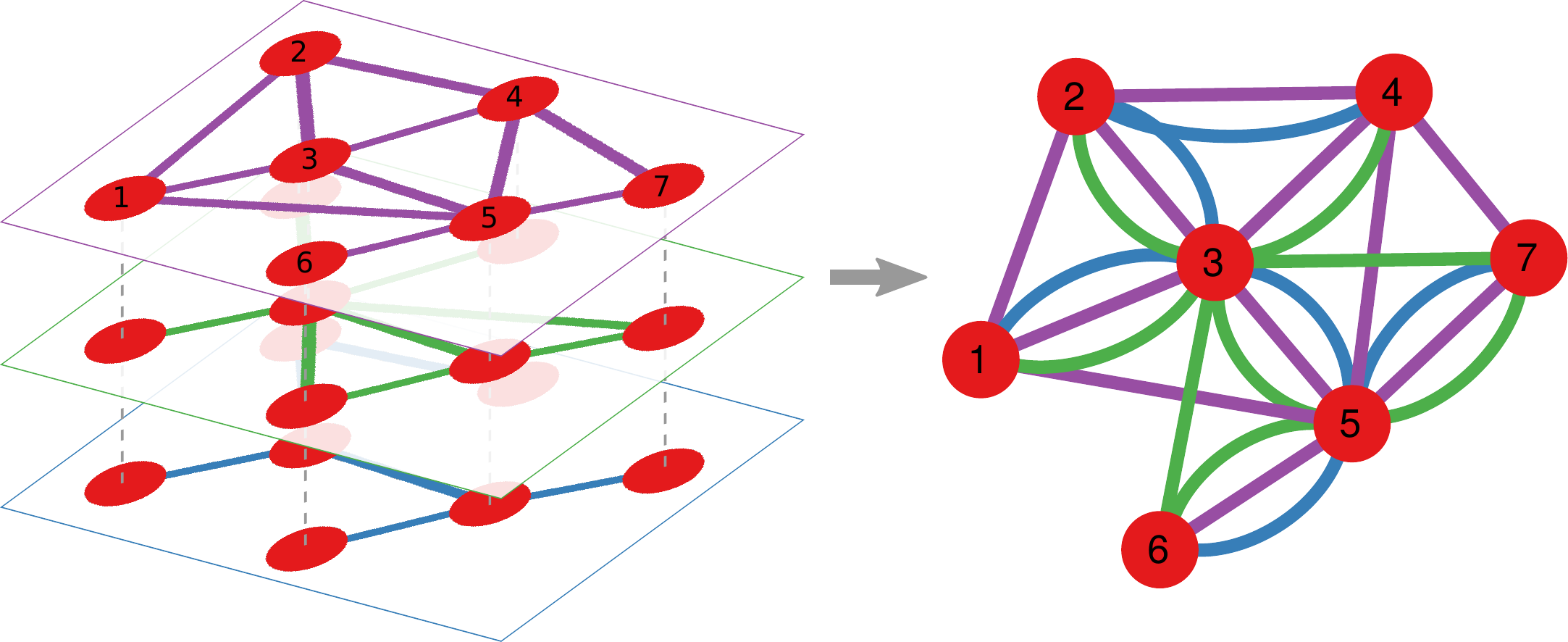}
\caption{Transforming a multilayer network with one-to-one couplings (left) into its corresponding labeled multigraph (right).}
\label{fig:model}
\end{figure}

\textbf{Directed Multiplex Network Model}. Our model is a multiplex network, which are labeled multigraphs. Multiplex networks are equivalent to multilayer networks with one-to-one inter-layer couplings, meaning that there is a one-to-one correspondence between nodes in different layers -- as Figure \ref{fig:model} shows.

Formally, a directed multiplex network is a quadruple $G = (V, L, E, A)$, where: $V$ is the set of nodes; $L$ is the set of link labels; $E$ is the set of multiplex links, i.e.~triples $(u,v,l)$, with $u,v, \in V$ and $l \in L$. The network is directed, thus $(u,v,l) \neq (v,u,l)$; $A$ is the set of categorical node attribute values -- which we use following previous works showing their usefulness in describing social status \cite{leskovec2010predicting}. Each node has a single attribute value $a \in A$.

\textbf{Multiplex Link Prediction}. Let us assume that $G_t$ represents the status of the multiplex graph $G$ at time $t$. Given two times, $t'$ and $t''$, with $t' < t''$, we expect $G_{t'} \neq G_{t''}$. Specifically, we assume that a certain set of links were added to $E_{t'}$. Our model could be extended in a straightforward way to cover the possibility of disappearing links \cite{noel2011unfriending} but for simplicity we follow traditional link prediction and only focus on the links that were added to $E_{t'}$. Specifically, we have a target set of links defined as $T = E_{t''}-E_{t'}$, the set of all links in $E_{t''}$ but not in $E_{t'}$. 

\begin{figure}
\centering
\includegraphics[width=.75\columnwidth]{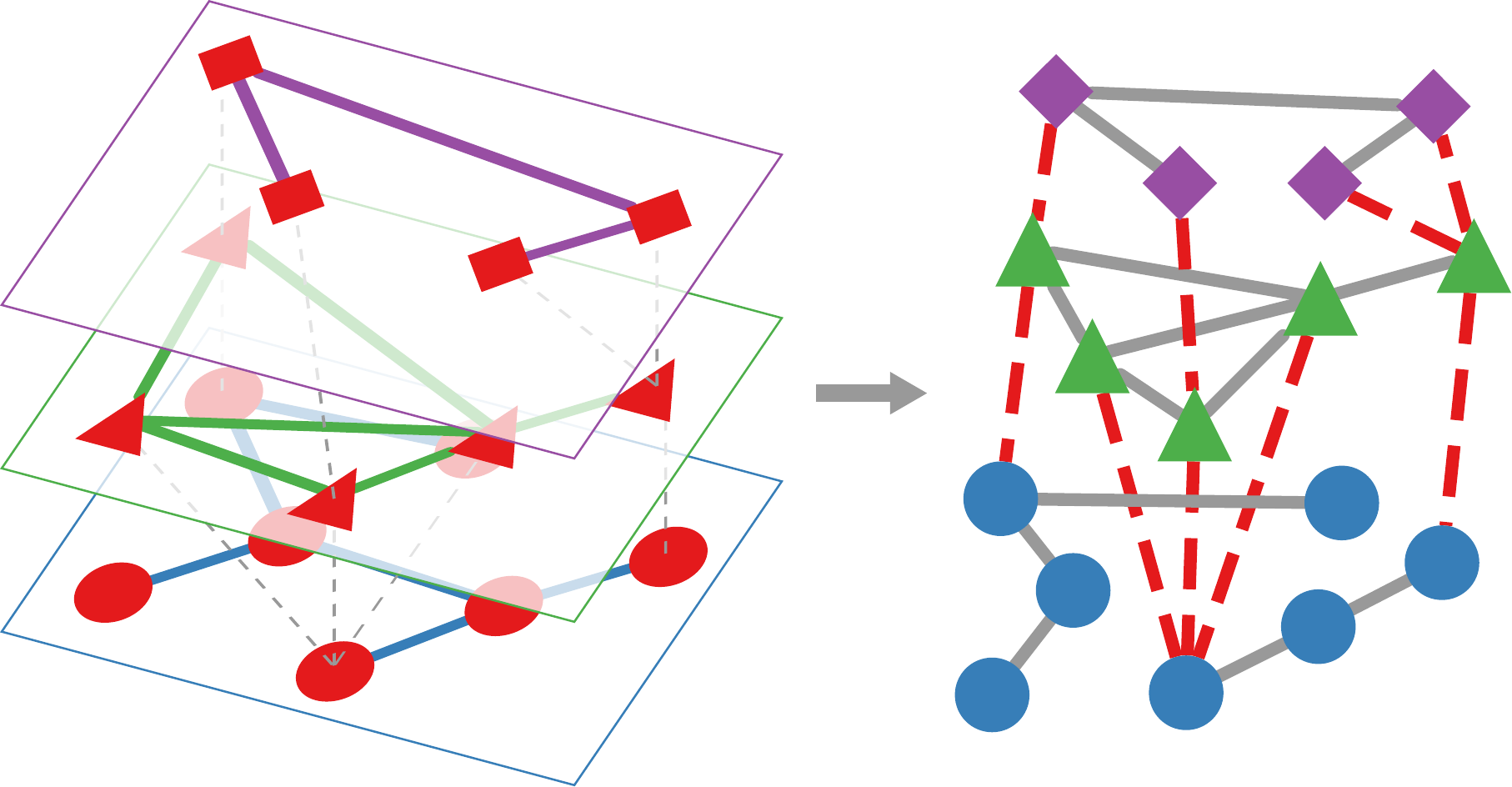}
\caption{Transforming a multilayer network with many-to-many couplings (left) into its corresponding labeled multigraph (right).}
\label{fig:many-to-many}
\end{figure}

\begin{figure*}
\centering
\includegraphics[width=.75\textwidth]{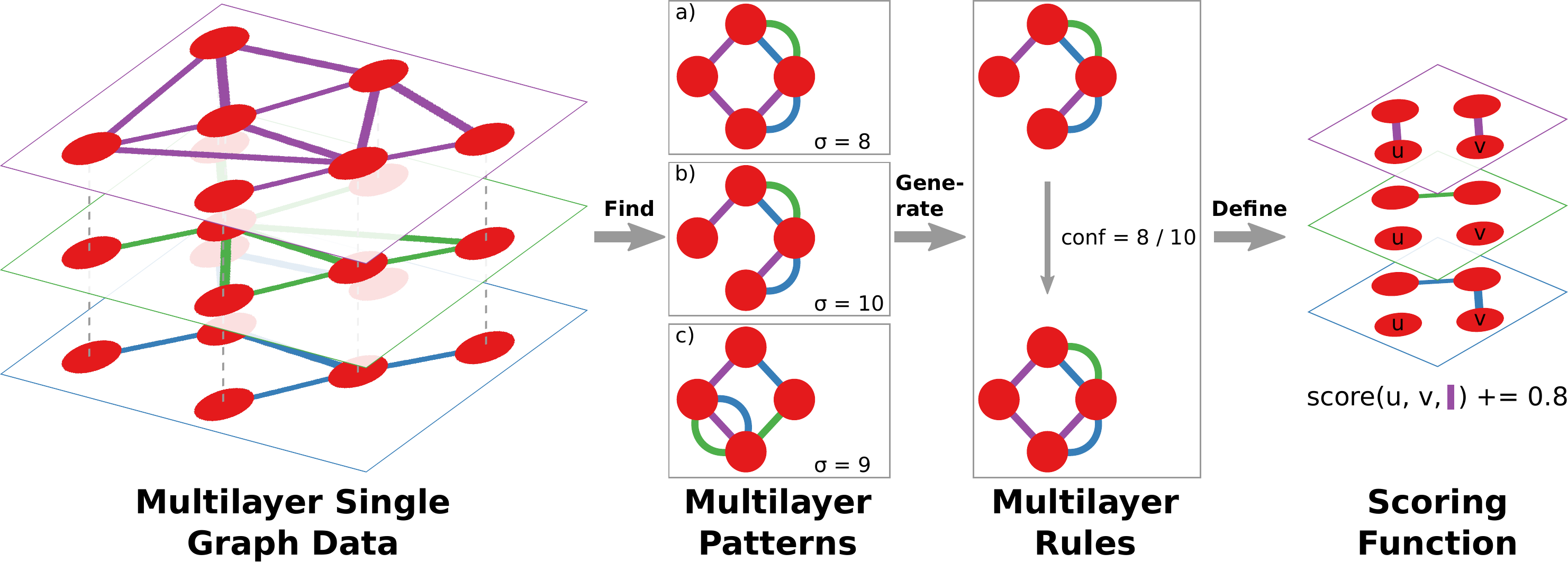}
\caption{The multilayer graph association rule mining framework.}
\label{fig:magma}
\end{figure*}

The link prediction problem is to estimate a $score(u,v)$ function for every missing link $(u,v) \not\in E_{t'}$. The $score(u,v)$ function should rate highly the missing links that are most likely to be part of $T$. In multiplex link prediction, the $score$ function takes an additional parameter: the link type $l$. Thus, our aim is to estimate $score(u,v,l)$, for every $(u,v,l) \not\in E_{t'}$. Since the multilayer network is directed, $score(u,v,l) \neq score(v,u,l)$.

\textbf{Extension to Many-to-Many Multilayer Networks}. While this paper focuses on multiplex networks, it is possible to use our framework to perform many-to-many multilayer network mining. This is achieved by adding a pre- and post-processor to transform the data.

Figure \ref{fig:many-to-many} illustrates the procedure. In the pre-processing phase, the multilayer graph is transformed in a simple graph with two edge types: links of type $1$ are inter-layer coupling, while nodes of type $2$ are regular intra-layer connections. Each node is labeled with the layer in which it appears. The post-processing phase undoes the pre-processing. Any frequent pattern found on the simple graph contains all the information to reconstruct the original multilayer pattern. Links of type $1$ connect the different identities of the same node across layers. Links of type $2$ are intra-layer connections and one can reconstruct to which layer they belong by looking at the layer information from the node label.

We choose to ignore this extension for the rest of the paper because it affects the interpretation of one of the parameters of our framework.

\section{Methods}\label{sec:meth}

\begin{figure}
\centering
\begin{subfigure}{.3\columnwidth}
\includegraphics[width=\textwidth]{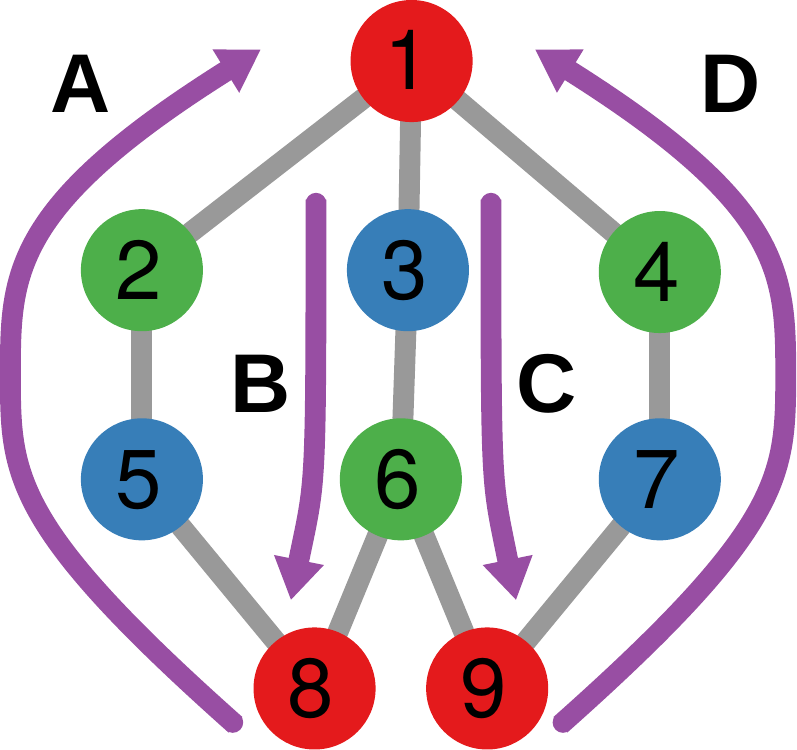}
\caption{}
\end{subfigure}\qquad
\begin{subfigure}{.04\columnwidth}
\includegraphics[width=\textwidth]{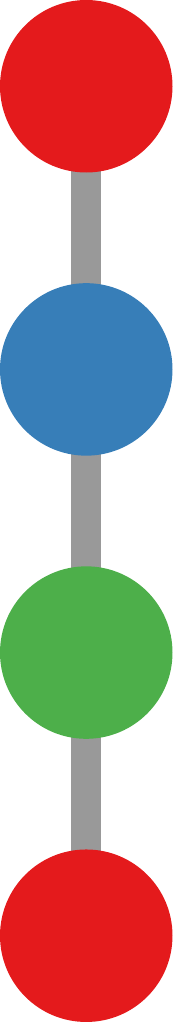}
\caption{}
\end{subfigure}\qquad
\begin{subfigure}{.4\columnwidth}
  \begin{tabular}{cccc|c}
    $A$ & $B$ & $C$ & $D$ & Count \\
    \hline
    $8$ & $1$ & $1$ & $9$ & $3$\\
    $5$ & $3$ & $3$ & $7$ & $3$\\
    $2$ & $6$ & $6$ & $4$ & $3$\\
    $1$ & $8$ & $9$ & $1$ & $3$\\
  \end{tabular}
\caption{}
\end{subfigure}
\caption{(a) The original graph, nodes labeled by their id and colored according to their label. Arrows indicate and label the occurrences of the motif in the graph. (b) The pattern. (c) The image table for the minimum image support definition, with the pattern's nodes as rows and all possible occurrences of the pattern as columns. Each cell records the node id we use for the mapping.}
\label{fig:mis}
\end{figure}

\textbf{The Framework}. Figure \ref{fig:magma} shows an overview of our framework. First, we find frequent multiplex graph patterns using \texttt{Moss}. \texttt{Moss} uses a minimum image based support definition to find frequent patterns in a single graph, and it accepts labeled multi-graphs as inputs. The support definition counts as the frequency of a pattern by estimating the number of different nodes in the original graph that can play a specific role in the pattern, and taking the minimum. Figure \ref{fig:mis} provides an example: there are four ways to map Figure \ref{fig:mis}(b) in Figure \ref{fig:mis}(a), but its support is three because we have to re-use the same node in the same role for some of these occurrences.

Since the link label represents its type, \texttt{GERM} cannot perform multiplex link prediction, as it already uses the link labels to determine the link's appearance time.

\texttt{Moss} requires two parameters: minimum support $\sigma$ and maximum pattern size $s$. The minimum support is the minimum number of occurrences of the pattern to be considered frequent and included in the results. The maximum pattern size is the maximum number of nodes in a pattern.

The second step is building the set $R$ of multiplex graph association rules. To keep complexity and potential overfitting under control, we decide to focus exclusively on rules which predict the appearance of a single new link.

In other words, we build a $p_1 \rightarrow p_2$ rule if pattern $p_2$ completely includes $p_1$, with a single additional edge. Moreover, we exclusively focus on connected rules, i.e. neither the antecedent nor the consequent have more than one connected component. Both $p_1$ and $p_2$ need to be frequent patterns, appearing more than $\sigma$ times in $G$. The single link differentiating $p_2$ from $p_1$ tells us the two nodes we expect to be connected and the link type. The weight of the rule is its confidence: the ratio of the support of the consequent over the support of the antecedent. Every time we encounter $p_1$ in $G$, we can identify the two nodes and the type of the missing link by looking at all its $p_2$ consequents in $R$.

In practice, $score(u, v, l)$ is the count of all rules saying $u$ should connect to $v$ in $l$, weighted by their confidence. There could be multiple weighting schemes -- simple count, lift, average confidence, etc. For simplicity, and to demonstrate performance and feasibility, we focus on a confidence-weighted score. In the experiment section, we discuss alternative scoring schemes equivalent to weighting by confidence, and additional ones which perform erratically.

\textbf{Predicting Old-New Links}. New links can attach to nodes that were not part of the network: $V_{t'} \subseteq V_{t''}$. There are two node classes in $V_{t''}$: ``old'' nodes which are nodes in $V_{t'}$, and ``new'' nodes from $V_{t''} - V_{t'}$. Each link in $E_{t''} - E_{t'}$ can belong in one of three categories: ``old-old'' links connect two old nodes, ``old-new'' links connect an old node with a new one, and ``new-new'' links connect two new nodes.

\begin{figure}
\centering
\includegraphics[width=.6\columnwidth]{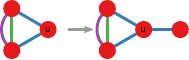}
\caption{A rule allowing us to predict old-new links. The link color represents its type.}
\label{fig:oldnew2}
\end{figure}

Traditional link prediction exclusively deals with scoring old-old links where all scoring functions need to be calculated on the topology of $G_{t'}$, thus nodes not in $G_{t'}$ cannot contribute to $score(u,v)$. On the contrary, here we are able to predict old-new links by exploiting the rules in which the consequent has one node more than the antecedent. To see how this is possible consider the rule in Figure \ref{fig:oldnew2}. Consequents are matched to antecedents if they contain them, minus one link. The new link is free to connect to an additional node, not necessarily to a node that was already part of the antecedent. Using the rule in the figure, we can predict that node $u$ will connect to a previously unobserved node.

\textbf{Parameter Choice}. Here we provide principled reasons on how to choose proper values for the minimum support $\sigma$ and the maximum pattern size $s$.

In data mining, the minimum support threshold is usually set as high as possible, because high support thresholds efficiently prune the search space, improving run times. However, setting the support too high leads to no patterns found. This also holds true here, with two additional considerations. First, increasing $\sigma$ decreases the number of found patterns. Up to a certain point, this \textit{improves} prediction performance because fewer patterns imply fewer and less specific rules, which in turn imply lower chances of overfitting. Second, we have a principled way to determine the hard upper limit of $\sigma$. This is the number of nodes of the smallest layer we want to predict. The minimum image support definition used by \texttt{Moss} is upper-bounded by the number of nodes in the network: for example, if layer $l$ has $|V_l| = 50$, setting $\sigma = 51$ guarantees there will be no pattern with a link in layer $l$. Thus, all new links in $l$ will receive a score of zero, with potentially devastating effects on the accuracy. Of course, setting $\sigma = |V_l|$ results in finding patterns that can only include a single edge from $l$, since any other more complex pattern in $l$ will have a support lower than $|V_l|$. 

For the maximum pattern size $s$ we suggest $s = 4$ for all but the smallest real world networks. There are already plenty of link prediction methods based on multiplex triangles, which is what the framework would reduce to if $s = 3$. However, the number of potential graph patterns increases exponentially with $s$. As we show in the parameter tuning section of the experiments, an increase in one unit of $s$ can lead to 10x more rules found.

\textbf{Computational Efficiency}. Our framework is experimental. Our aim is to show that multiplex association rules provide a significant prediction advantage in principle, providing arguments for their use in link prediction. Therefore the current implementation is deliberately neglecting runtime efficiency, which we leave for future work to optimize.

Notwithstanding its prototypical nature, this approach is by design more memory efficient than the competitors: instead of giving a score to all unconnected node pairs, which grow quadratically with the number of nodes, graph association rules will only score a generally much more limited number of node pairs that can appear in a rule. 

\section{Experiments}
In our experiments, we first present the baseline algorithms and the data we use to make the comparison. Then we explore the effect of parameter choices. We move on comparing our performance with a set of baselines and an ideal ensemble classifier. Finally, we discuss interesting patterns we can extract showing insights from motifs that go beyond three nodes. To save space, we label our framework as \texttt{MAGMA} (Multiplex Association Graph Mining Analysis).

\subsection{Setup}\label{sec:exp-setup}
\subsubsection{Datasets.}

\begin{table}[]
\centering
\begin{tabular}{l|rrrrr}
Network & $|V|$ & $|E|$ & $|L|$ & Dir & Dyn\\
\hline
Aarhus & 61 & 620 & 5 & N & N\\
Physicians & 241 & 1,551 & 3 & Y & N\\
CElegans & 279 & 5,863 & 3 & Y & N\\
Pardus & 6,373 & 78,661 & 3 & Y & Y\\
Synthetic & 200 & 2,170 & 4 & N & N\\
\end{tabular}
\caption{Basic statistics of the datasets: $|V|$, number of nodes; $|E|$, number of links; $|L|$, number of layers; Dir, whether the network is directed; Dyn, whether the network has temporal information.}
\label{tab:data}
\end{table}

Aarhus \cite{magnani2013combinatorial} records interactions in the CS department of Aarhus University. Employees can establish five different types of relations: coauthorship, lunch, collaboration, Facebook friendship, and leisure time. This is a static undirected network.

Physicians \cite{coleman1957diffusion} tracks relations between physicians asking three questions. Each physicians reports with whom they: ask advice, discuss cases, and/or have a friendship relations. Each question generates a link type in the network. This is a directed network.

CElegans \cite{chen2006wiring} is the neurological structure of the C. Elegans worm. There are three types of connections, each corresponding to a different link type: electric, chemical monadic, and chemical polyadic.

Pardus \cite{szell2010multirelational,szell2010msd} includes relations between players from an online game.\footnote{\url{https://www.pardus.at/}} Players can be each other's friends or enemies, and can attack each other. This generates three layers, one positive (friendship) while the others (enemies and attacks) are negative. This is a temporal directed network. We use the network on day 300 as training set, and the network observed 100 days later as the test set.

We also generate synthetic data from four LFR benchmarks \cite{lancichinetti2008benchmark}, one per link type. All parameters for the benchmark are the same across layers except the number of nodes. The layers have $200$, $150$, $100$, and $50$ nodes, to illustrate the relationship between layer size, $\sigma$ parameter, and classifier accuracy.

Table \ref{tab:data} reports basic statistics of our datasets. The reported sizes (number of nodes $|V|$, and edges $|E|$) of the datasets are the unions of their training and test sets (both the number of nodes and links might increase from training to training+test, as new nodes might be introduced). All datasets except Pardus come from the CoMuNe project\footnote{\url{https://comunelab.fbk.eu/data.php}} \cite{de2013mathematical}. We remove all self loops. 

Some datasets have temporal information and some do not. For the datasets without temporal information, we perform the link prediction task using ten-fold cross validation as the split between training and test. We build each test set by randomly drawing 10\% of the edges, which means that it might contain nodes that are not in the training fold -- if we picked all of their edges. For the dataset with temporal information, we collect data until time $t$ for the training data, and we use data starting from time $t$ until $t + \delta$ for the test.

\subsubsection{Baseline algorithms.}\label{sec:exp-setup-base}
Here we briefly present the state of the art of multiplex link prediction.

Sharma \cite{sharma2016efficient} calculates the likelihood of having a link of type $l_1$ given that the nodes are connected by link type $l_2$: $p_{l_2,l_1}$. Then, 

$$score(u,v,l_1) = \sum \limits_{l_2 \in L} p_{l_2,l_1} \delta_{u,v,l_2},$$

with $\delta_{u,v,l_2}$ being equal to $1$ if nodes $u$ and $v$ are connected in $l_2$, $0$ otherwise. The downside is that every node pair not connected in any layer will get a score of zero. While this makes it the most memory efficient approach by dramatically reducing output size, it also makes it miss all connections between previously completely disconnected nodes, which routinely happen in real world networks.

Pujari \cite{pujari2015link} takes a collection of classical link prediction scores (Common Neighbor, Adamic-Adar, etc.) for each link type separately as input features for a decision tree. It adds multiplex features such as the score average and entropy across layers. A disadvantage is a lack of feature for pairwise link type interactions, only for the overall interaction between all link types pairs. A related method \cite{hajibagheri2016holistic} adds temporal information, but reduces to the Pujari method for static networks, thus for our purposes they are equivalent.

Jalili \cite{jalili2017link} builds a metagraph by performing community discovery on each link type separately using Infomap \cite{rosvall2008maps}, then it counts the number of simple metapaths of length $1$, $2$, and $3$ that lead from node $u$ to node $v$ either starting or ending in layer $l$. It generates six features as the input of an SVM with a Gaussian kernel. Paths cannot contain cycles -- however, it is possible to calculate them by multiplying the adjacency matrix with itself and removing the diagonal, since the paths are capped to be of length 3.

Hristova \cite{hristova2016multilayer} calculates a series of classical scores per link type. It then generates multiplex features by aggregating these scores, and feeds them to a Random Forest classifier. The original paper defines a number of features that are inapplicable here because they are tailored for special geotemporal data (Twitter and Foursquare). They also define two multiplex aggregations, which they call ``global'' and ``core''. Here we only use the global one, as the core aggregation is too restrictive and leads to a too sparse output.

De Bacco \cite{de2017community} defines a multilayer mixed-membership stochastic blockmodel \cite{airoldi2008mixed} by assuming that nodes belong to the same groups across layers -- a more relaxed version has also been recently proposed \cite{roxana2019edge}. The group-group affinity is different in each layer, allowing for pairs of layers to be correlated, anti-correlated, or independent from each other. It then finds the best node-node and group-group connection probabilities via the expectation maximization algorithm which serve as the scores for the link prediction task.

\subsubsection{Parameter Tuning}
We use the Synthetic dataset to study the sensitivity of our framework to its two parameters: minimum support threshold $\sigma$ and output pattern size $s$.

\begin{figure}
\centering
\includegraphics{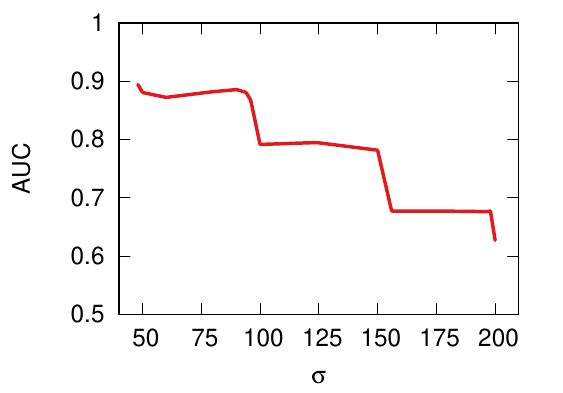}
\caption{AUC score (y axis) at different levels of support parameter (x axis) for the Synthetic dataset.}
\label{fig:auc-synth}
\end{figure}

\begin{figure*}
\centering
\begin{subfigure}[t]{.23\textwidth}
\includegraphics[width=\textwidth]{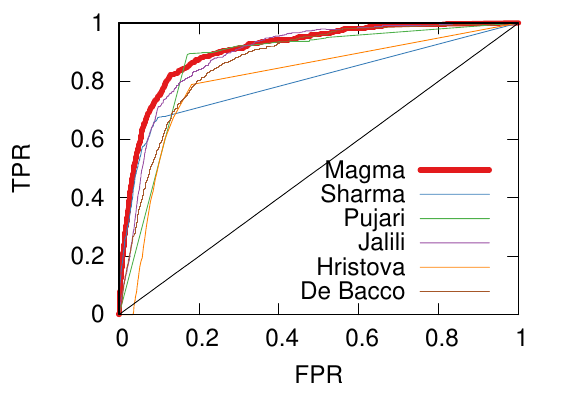}
\caption{Aarhus}
\end{subfigure}
\enskip
\begin{subfigure}[t]{.23\textwidth}
\includegraphics[width=\textwidth]{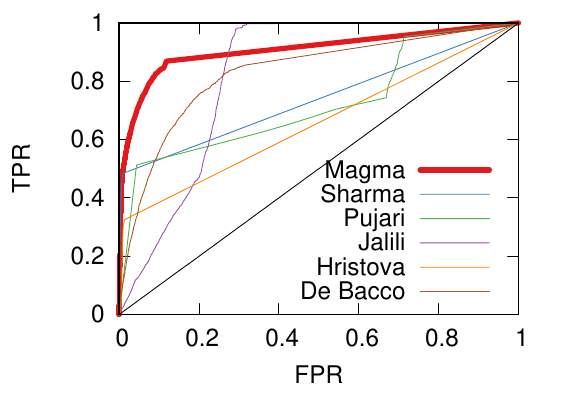}
\caption{Physicians}
\end{subfigure}
\enskip
\begin{subfigure}[t]{.23\textwidth}
\includegraphics[width=\textwidth]{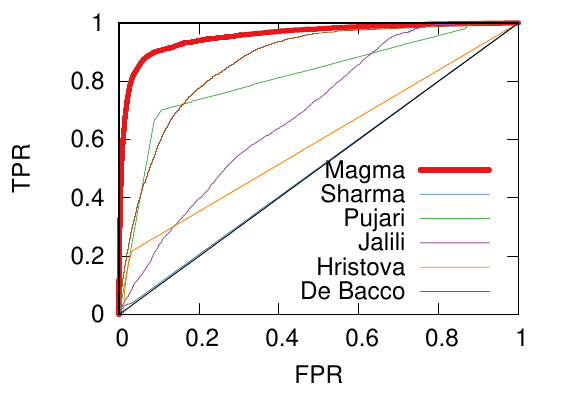}
\caption{CElegans}
\end{subfigure}
\enskip
\begin{subfigure}[t]{.23\textwidth}
\includegraphics[width=\textwidth]{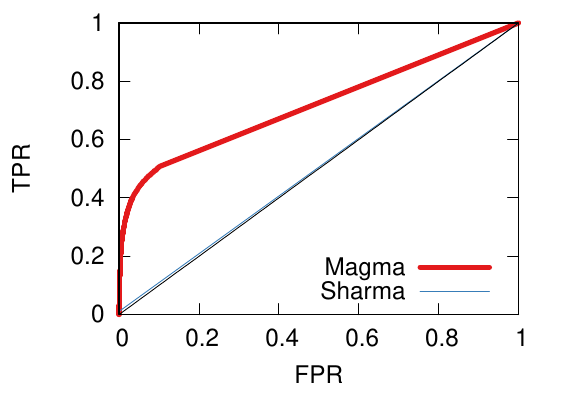}
\caption{Pardus}
\end{subfigure}
\caption{ROC curves for all methods that managed to finish and all empirical datasets.}
\label{fig:roc-aarhus}
\end{figure*}

Figure \ref{fig:auc-synth} shows the AUC performance at different levels of support. We make two observations.

First, there are dips as we approach critical values of the support threshold $\sigma$ ($50$, $100$, $150$, and $200$): the number of nodes of the four layers in the Synthetic network. If we set a support threshold higher than the number of nodes in a layer, we become unable to provide a prediction for links of that type, with evident detrimental effects to the performance.

Second, the performance reaches different local optima in each step, rather than having the local optima at the lowest values of $\sigma$ as would be expected given the highest number of patterns found for lower support threshold values. Higher $\sigma$ values allow to focus on fewer and more general rules. This helps to avoid overfitting, increasing the performance.

% \begin{table}
% \centering
% \begin{tabular}{l|rr}
% & $s = 3$ & $s = 4$ \\
% \hline
% \# Patterns & 14 & 83\\
% \# Rules & 17 & 177\\
% AUC & 0.5035 & 0.8724\\
% \end{tabular}
% \caption{The size and performance (rows) of the rule set on the Synthetic dataset, depending on $s$, the pattern size parameter (columns).}
% \label{tab:auc-psize}
% \end{table}

The effect of the maximum pattern size $s$ parameter is as follows: If we set $s = 3$ we only consider multiplex triangles. As a result, we find few patterns and only 17 rules, and the performance is almost nil (AUC $\approx 0.504$). On the other hand, allowing $s = 4$ nodes shows dramatic improvements in the AUC performance (AUC $\approx 0.872$) due to the increased rule set size of 177.

From this analysis we can conclude that: (i) the support parameter should be set as high as possible -- to avoid overfitting -- with the hard maximum being the number of nodes in the smallest layer to be considered -- to allow the miner to return rules involving that link type. Also (ii) the ability of extracting patterns involving more than three nodes is the key to the graph association rules' performance.

\subsection{Performance}\label{sec:exp-base}

\subsubsection{Multiplex}
We test the performance of multiplex graph association rules against the state of the art on all networks (except Pardus because most baseline approaches require too much time or memory to handle it).

We use the standard approach of building a ROC curve and calculating the area under the curve (AUC) as evaluation. Figure \ref{fig:roc-aarhus} shows the ROC curves for all the methods on all datasets. \texttt{MAGMA} consistently outperforms the state of the art at almost all levels of confidence, with rare exceptions. One exception is on the Aarhus data, where Pujari peaks higher. However, Pujari gives the same high score to many links, thus making it difficult to tune the false positive rate, which might be problematic in cases when false positives are more costly. The second exception is Jalili in the Physicians data. It arises mainly due to our choice of $\sigma$ which is slightly too strict to return all relevant patterns.

\begin{table}
\small
\centering
\begin{tabular}{l|rrrr}
Method & Aarhus & Physicians & CElegans & Pardus\\
\hline
\texttt{MAGMA} & {\bf 0.909} & {\bf 0.904} & {\bf 0.957} & {\bf 0.719}\\
Sharma & 0.800 & 0.738 & 0.504 & 0.506\\
Pujari & 0.866 & 0.694 & 0.815 & OOM\\
Jalili & 0.892 & 0.823 & 0.689 & OOM\\
Hristova & 0.806 & 0.655 & 0.596 & OOM\\
De Bacco & 0.869 & 0.820 & 0.867 & OOM\\
\hline
Ensemble Base & 0.921 & 0.933 & 0.933 & OOM\\
Ensemble Over & 0.933 & 0.951 & 0.966 & OOM\\
\end{tabular}
\caption{AUC of the ROC curves from Figure \ref{fig:roc-aarhus}. Highest performance values among the non-ensemble methods in bold. OOM: Out Of Memory.}
\label{tab:aucs}
\end{table}

Notwithstanding these exceptions, \texttt{MAGMA}'s AUC is the highest of all methods tested. Table \ref{tab:aucs} shows all AUC values. The improvement over the second best method spans from a minimum of 1.6\% to a maximum of 8.9\%, ignoring Pardus. The second best performing algorithm is different across datasets (Jalili in Aarhus and Physicians, De Bacco in CElegans), further proving the consistency of \texttt{MAGMA}.

\begin{figure*}
\centering
\begin{subfigure}[t]{.23\textwidth}
\includegraphics[width=\textwidth]{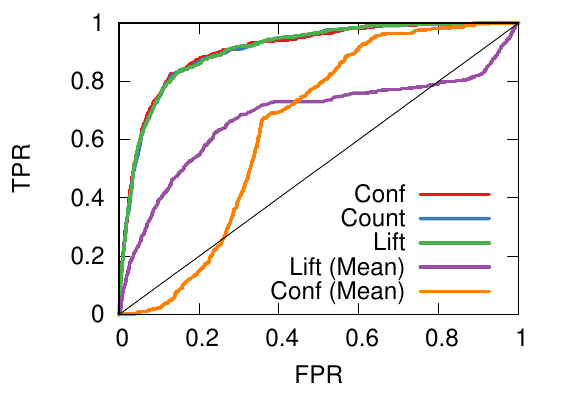}
\caption{Aarhus}
\end{subfigure}
\enskip
\begin{subfigure}[t]{.23\textwidth}
\includegraphics[width=\textwidth]{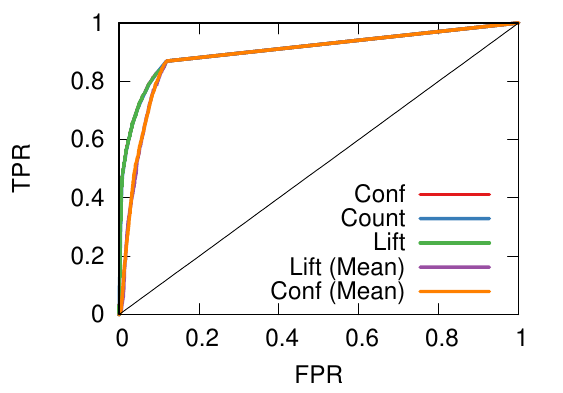}
\caption{Physicians}
\end{subfigure}
\enskip
\begin{subfigure}[t]{.23\textwidth}
\includegraphics[width=\textwidth]{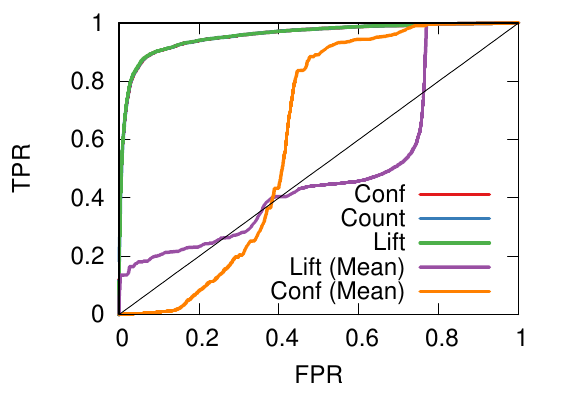}
\caption{CElegans}
\end{subfigure}
\caption{The comparison between different scoring functions for \texttt{MAGMA} for different datasets.}
\label{fig:magma-scoring}
\end{figure*}

In our framework, we decide to use as a scoring function the sum of the confidence of all rules that apply for a specific link (Conf). There are alternative weighting schemes. Namely we could: count the number of rules without weights (Count), count weighting by the lift (Lift) -- i.e. the overexpression with respect to change --, average the confidence values (Conf Mean), or average the lift values (Lift Mean) of all applicable rules for the link we are predicting. Figure \ref{fig:magma-scoring} shows how these choices impact \texttt{MAGMA}'s performance on the Aarhus, Physicians and CElegans datasets. The curves for Count, Conf, and Lift are equivalent because most of the information is contained in the number of rules that apply: confidence and lift only provide corrections. Both of the averaging rules perform poorly and should thus be avoided. The inverse of the lift average could be considered, as we do in the Aarhus dataset, but the interpretability of this measure would be questionable.

\begin{figure*}
\centering
\begin{subfigure}[t]{.23\textwidth}
\includegraphics[width=\textwidth]{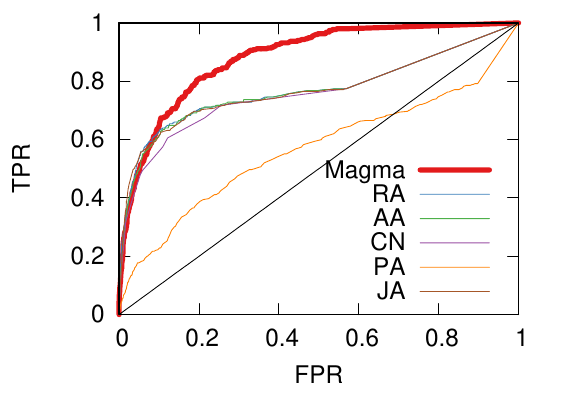}
\caption{Aarhus}
\end{subfigure}
\enskip
\begin{subfigure}[t]{.23\textwidth}
\includegraphics[width=\textwidth]{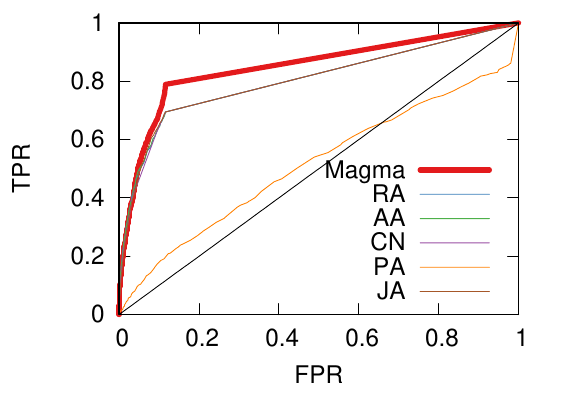}
\caption{Physicians}
\end{subfigure}
\enskip
\begin{subfigure}[t]{.23\textwidth}
\includegraphics[width=\textwidth]{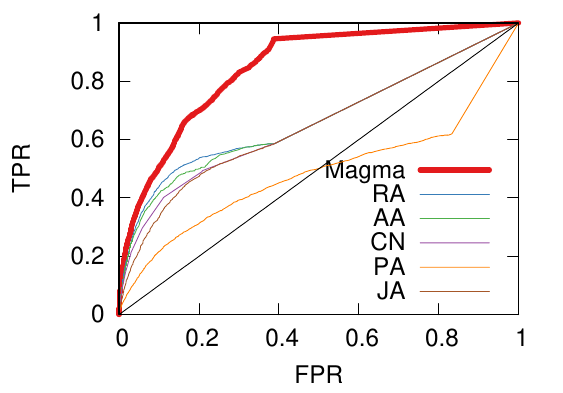}
\caption{CElegans}
\end{subfigure}
\caption{ROC curves for all single layer methods on collapsed single layer versions for different datasets.}
\label{fig:roc-aarhus-sl}
\end{figure*}

\subsubsection{Single Layer}
\texttt{MAGMA} is a multiplex version of \texttt{GERM}, but it can still be applied to single layer networks. We build single layer versions of the above four datasets by collapsing the multiplex information: we connect nodes if they share a link, regardless of their type.

\begin{table}
\centering
\begin{tabular}{l|rrr}
Method & Aarhus & Physicians & CElegans\\
\hline
\texttt{MAGMA} & {\bf 0.882} & {\bf 0.849} & {\bf 0.845}\\
RA & 0.772 & 0.805 & 0.671\\
AA & 0.770 & 0.805 & 0.665\\
CN & 0.759 & 0.803 & 0.655\\
PA & 0.567 & 0.515 & 0.488\\
JA & 0.771 & 0.804 & 0.648\\
\hline
Ensemble Base & 0.815 & 0.800 & 0.752\\
Ensemble Over & 0.916 & 0.883 & 0.878\\
\end{tabular}
\caption{AUC of the ROC curves from Figure \ref{fig:roc-aarhus-sl}.}
\label{tab:aucs-sl}
\end{table}

We only test \texttt{MAGMA} against classical single layer link prediction techniques, since here we are interested in reproducing \texttt{GERM}'s performance in single layer cases. This means we test against Resource Allocation (RA), Adamic-Adar (AA), Common Neighbor (CN), Preferential Attachment (PA), and JAccard (JA). Figure \ref{fig:roc-aarhus-sl} shows the ROC curves for all tests. Table \ref{tab:aucs-sl} shows the AUC values. We see a strong performance of \texttt{MAGMA} when comparing to classical link prediction approaches, especially in CElegans.

We are aware that these methods are by now outdated, but in this secondary test we merely want to reproduce \texttt{GERM}, and we thus compare \texttt{MAGMA} against what was available at the time of the development of \texttt{GERM}. This test shows that \texttt{GERM} -- a patented, closed source software not available to use -- can be replaced by \texttt{MAGMA}.

\subsubsection{Pardus}
The Pardus dataset deserves a detailed discussion: most multiplex link predictors fail on this dataset due to their memory requirements. Sharma handles the network, but only because it exclusively looks at pairs of nodes already connected in at least one layer. In Pardus this works poorly: friends cannot be enemies, so these two layers do not share links, forcing Sharma to search for impossible links.

\begin{figure}[t!]
\centering
\includegraphics[width=.5\columnwidth]{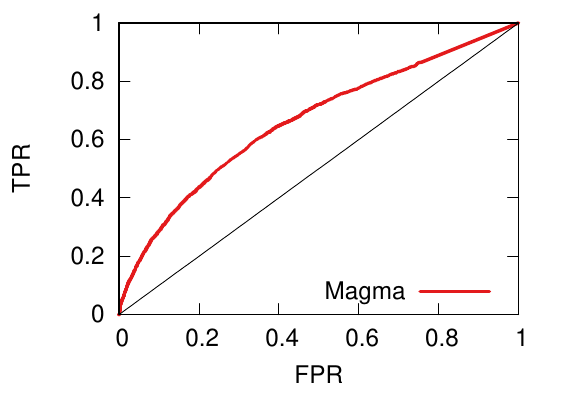}
\caption{\texttt{MAGMA}'s ROC curve for Pardus focusing only on the old-new link types.}
\label{fig:roc-pardus-onlynew}
\end{figure}

Next we use the Pardus dataset to test \texttt{MAGMA}'s added ability of predicting old-new links. In Figure \ref{fig:roc-pardus-onlynew} we replicated Figure \ref{fig:roc-aarhus}, but only considering old-new links. The performance understandably drops -- the AUC in this case is 0.663 -- because the problem is harder: we need to predict 1) that a new node will appear, 2) to which old node it will connect, and 3) of which type. \texttt{MAGMA}'s ability to make a better-than-chance prediction here is in itself remarkable, even more so considering that none of the other tested link prediction methods can make any guess.

\begin{figure*}
\centering
\begin{subfigure}[t]{.23\textwidth}
\includegraphics[width=\textwidth]{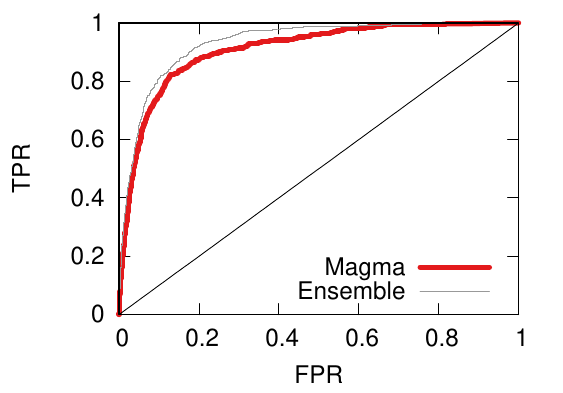}
\caption{Aarhus -- Multiplex}
\end{subfigure}
\enskip
\begin{subfigure}[t]{.23\textwidth}
\includegraphics[width=\textwidth]{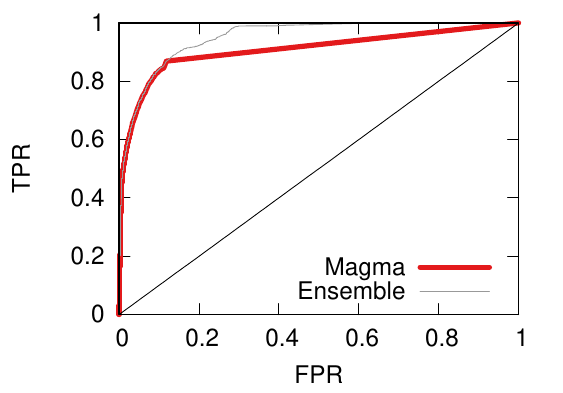}
\caption{Physicians -- Multiplex}
\end{subfigure}
\enskip
\begin{subfigure}[t]{.23\textwidth}
\includegraphics[width=\textwidth]{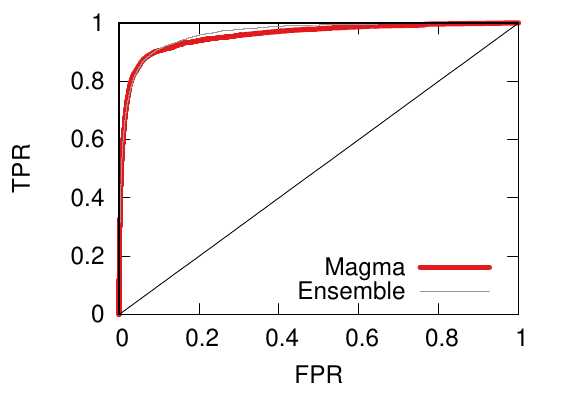}
\caption{CElegans -- Multiplex}
\end{subfigure}\\
\begin{subfigure}[t]{.23\textwidth}
\includegraphics[width=\textwidth]{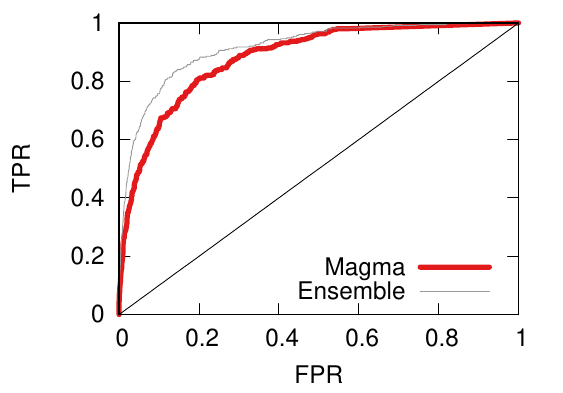}
\caption{Aarhus -- Single layer}
\end{subfigure}
\enskip
\begin{subfigure}[t]{.23\textwidth}
\includegraphics[width=\textwidth]{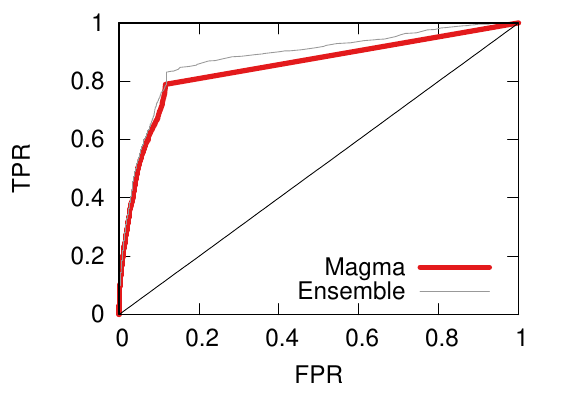}
\caption{Physicians -- Single layer}
\end{subfigure}
\enskip
\begin{subfigure}[t]{.23\textwidth}
\includegraphics[width=\textwidth]{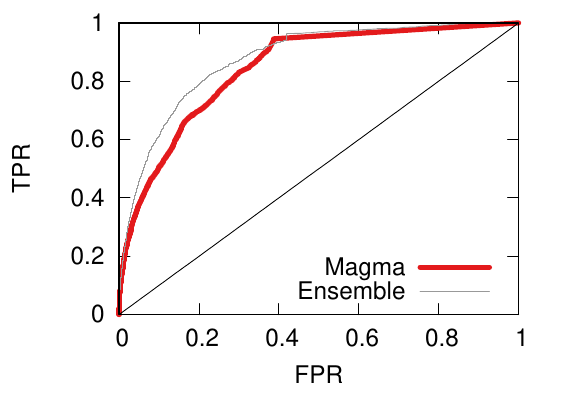}
\caption{CElegans -- Single layer}
\end{subfigure}
\caption{Comparing the ROC curves of \texttt{MAGMA} (red) with the ensemble classifier (gray) on all datasets except Pardus.}
\label{fig:roc-aarhus-ensemble}
\end{figure*}

\subsubsection{Ensemble}
We combine all classifiers tested so far into an ensemble classifier incorporating all scores. Ensemble classifiers use all available information providing an upper bound for the performance. It is useful to assess how much our method could be improved by adding more information.

Our ensemble classifier works in two steps. First, it normalizes the scores of the methods so that their average equals zero and their standard deviation equals one. This way, all classifier scores are on the same scale. Second, it searches via simulated annealing for the best weighting score, i.e. the one maximizing prediction quality, by multiplying each predictor score by a weight.

Figure \ref{fig:roc-aarhus-ensemble} shows the performance of the ensemble classifier on all datasets except Pardus, compared with \texttt{MAGMA} by itself. Tables \ref{tab:aucs} and \ref{tab:aucs-sl} report the ensemble's AUC scores.

First, the difference between the ensemble and \texttt{MAGMA} tends to be in the same range or lower than the difference between \texttt{MAGMA} and the second best performing classifier. This means that \texttt{MAGMA} is closer to an ideal classifier than any alternative is to \texttt{MAGMA}. In the multiplex case we have (ensemble vs \texttt{MAGMA} first, compared to \texttt{MAGMA} vs second best): 2.5\%-1.6\% in Aarhus, 4.7\%-8.2\% in Physicians, and 0.8\%-8.9\% in CElegans. In the single layer case the difference is even higher, due to correlations between baselines: 3.5\%-11\% in Aarhus, 3.4\%-4.5\% in Physicians, and 3.3\%-17\% in CElegans.

Second, the ensemble is overfitted, as the simulated annealing step cannot be performed with a training-test split, and thus unfairly boosts the ensemble's performance. For this reason we label it ``Ensemble Over'' in Table \ref{tab:aucs}. Without the simulated annealing step, equally weighting all methods, the ensemble has a lower AUC -- ``Ensemble Base'' in Table \ref{tab:aucs}. This AUC tends to be closer to \texttt{MAGMA}'s performance and, in some cases -- like in CElegans for the multiplex case, or all cases in the single layer link prediction --, lower.

We conclude that, even pooling all information available from all proposed methods, \texttt{MAGMA} is close to optimal performance. There is little information that can be added to \texttt{MAGMA} by using the alternative state of the art methods.

\subsection{Case Study: Pardus}\label{sec:exp-case}
Here we discuss a few of the significant patterns we find in the Pardus network\footnote{All patterns discussed here have lift $> 1$, which implies that they are overexpressed against null expectation and, thus, significant.}, which allow us to explore the data from two perspectives. First, we investigate how patterns with four nodes provide possible extensions/corrections to social balance theory. Then, we investigate dynamics in the Pardus game, enlightening us on the thought processes some players have when forming social networks online.

\subsubsection{Long Range Social Balance}\label{sec:exp-case-balance}
Social balance theory \cite{heider1958psychology,antal2005dynamics,leskovec2010predicting,szell2010multirelational,kirkley2019balance} looks at triangles in signed networks to predict the sign of new links. The expectation is that triangles will be balanced: friends of friends will be friends, while a friend's enemy is an enemy.

\begin{figure}
\centering
\includegraphics[width=.41\columnwidth]{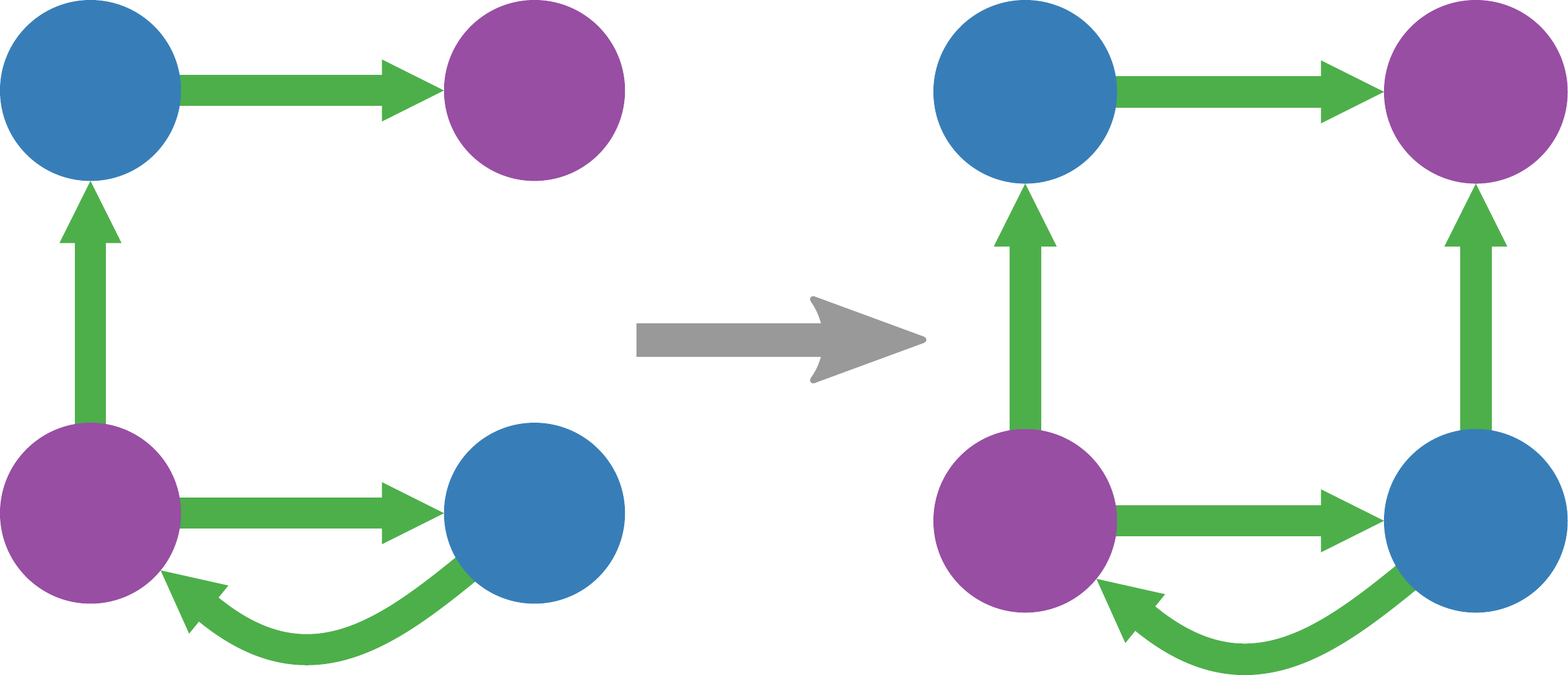}\qquad \qquad
\includegraphics[width=.41\columnwidth]{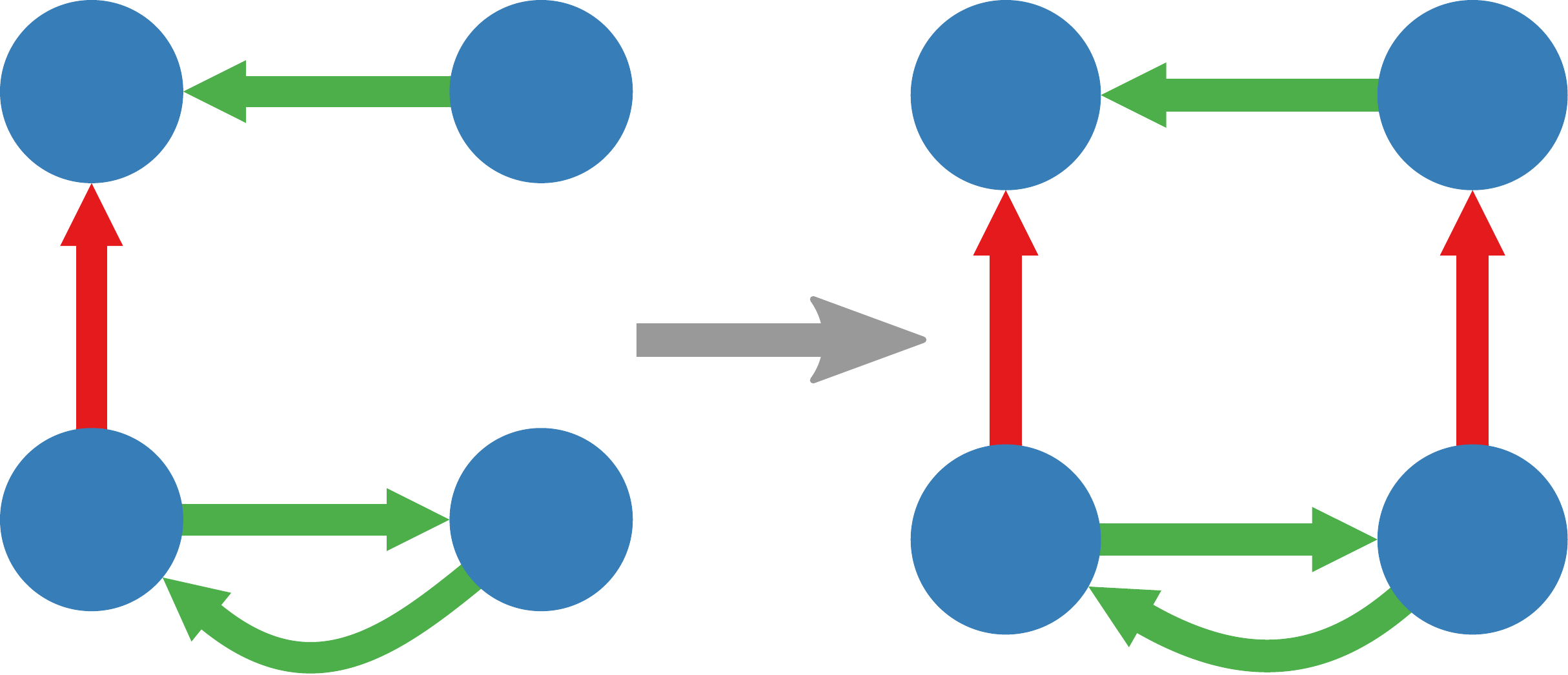}
\caption{Balanced friend (left) and enemy (right) 4-closure in Pardus. Node colors: purple = power player; blue = non-power player. Link colors: green = friendship; red = enmity.}
\label{fig:pardus1}
\end{figure}

Social balance theory is limited by its focus on triangles. Here we have no such limitation, so we can explore how people interact in a signed network in groups of four. Figure \ref{fig:pardus1} shows some examples of significant patterns conforming to the expectations of balance extended to four nodes. In Figure \ref{fig:pardus1} (left) we see a group of friends getting more closely knit: two people become friends because they each have friends which are friends to each other. Figure \ref{fig:pardus1} (right) shows a complementary pattern: a friend marking as enemy the friend of his friend's enemy.

\begin{figure}
\centering
\includegraphics[width=.41\columnwidth]{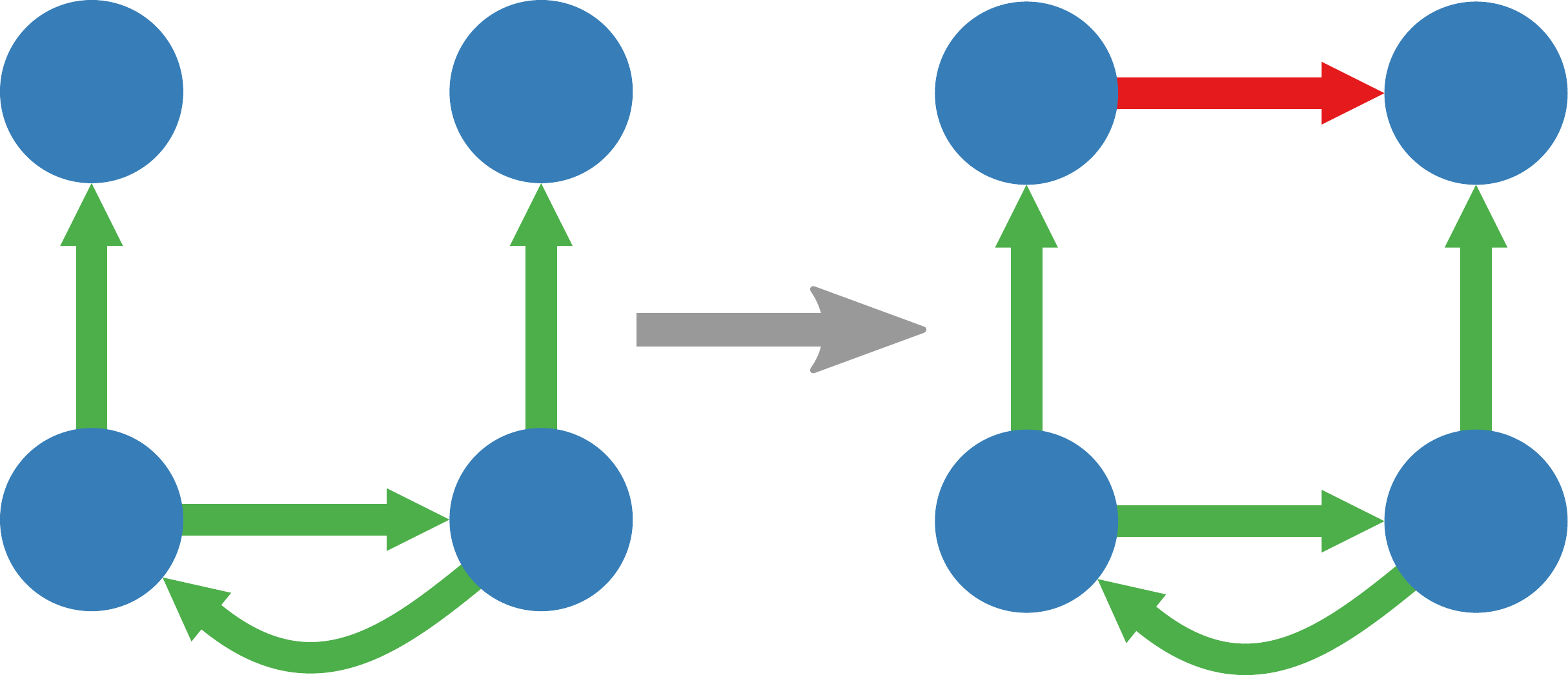}\qquad \qquad
\includegraphics[width=.41\columnwidth]{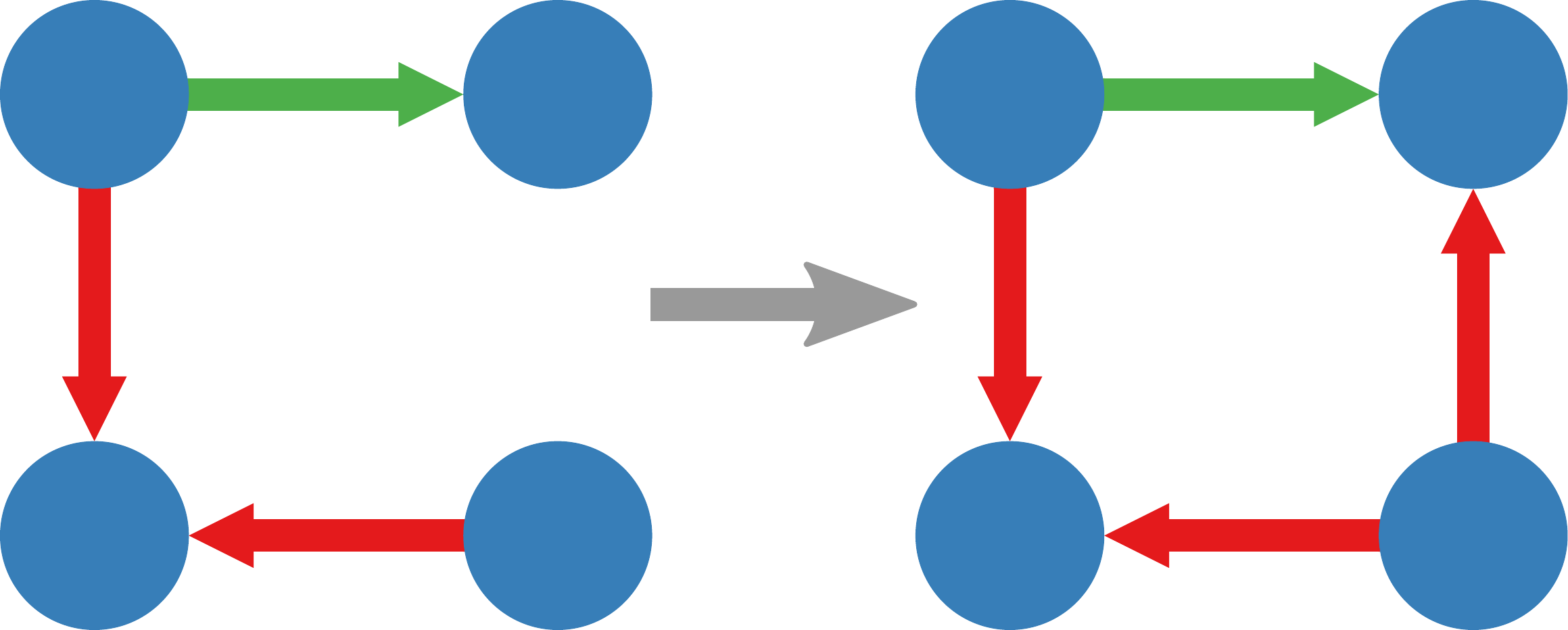}
\caption{Unbalanced friend (left) and enemy (right) 4-closure in Pardus. Node and link colors as in Figure \ref{fig:pardus1}.}
\label{fig:pardus2}
\end{figure}

However, we also find some significant patterns defying what we would expect in social balance. In Figure \ref{fig:pardus2} (left), we would expect the closing link to be positive, completing a tribe. Yet, the player is enemy to a friend's friend. Similarly, we would expect a positive completing in Figure \ref{fig:pardus2} (right): since the node completing the square has a common enemy with a friend of the target, they should befriend the target. Yet, they are an enemy to them.

These suggestive patterns suggest a hypothesis. In online environments, where the search space is too large, it might be too hard to assess friendships and alliances. Thus, we expect social balance rules to be routinely broken when considering structures of higher order than triangles.

\subsubsection{Game Dynamics}\label{sec:exp-case-pardus}
Apart from social balance, we present two more examples of how the graph association rule approach allows us to extend previous insights on the social behavior of Pardus players \cite{szell2010multirelational,szell2010msd}. 

\begin{figure}
\centering
\includegraphics[width=.41\columnwidth]{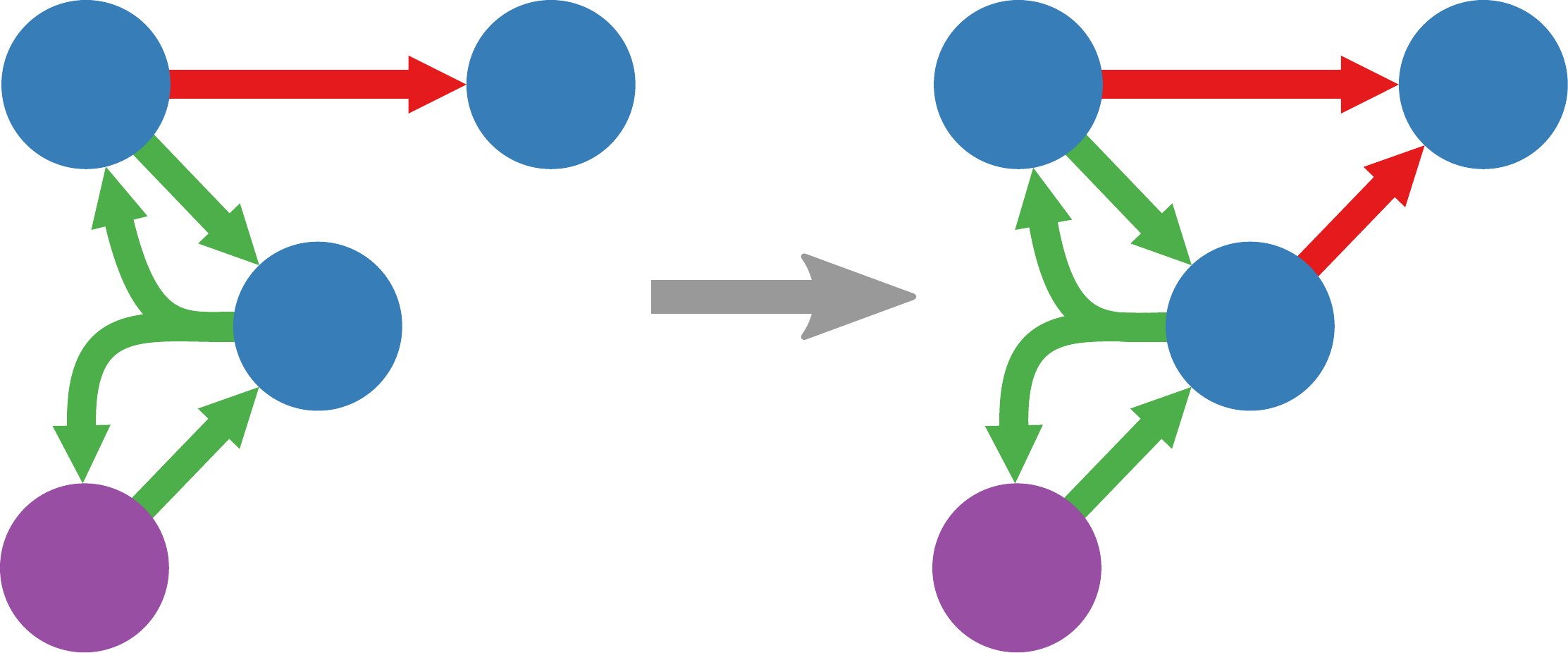}\qquad \qquad
\includegraphics[width=.41\columnwidth]{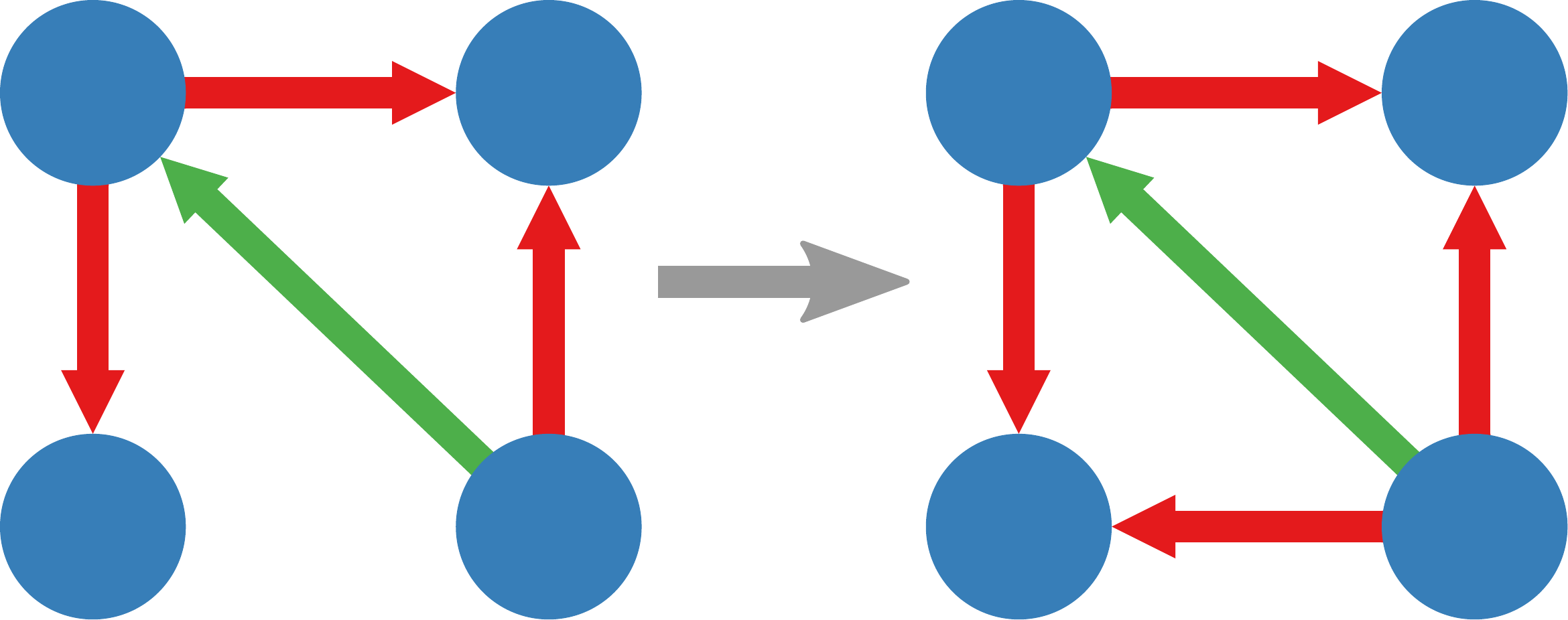}
\caption{(Left) Strong backing enemy closure in Pardus. (Right) Ganging up on common enemies in Pardus. Node and link colors as in Figure \ref{fig:pardus1}.}
\label{fig:pardus3}
\end{figure}

The first pattern we report in Figure \ref{fig:pardus3} (left) is describing a balanced closure of a player marking her friend's enemy as enemy. It extends the friend dyad with a power player (a player with more game experience, purple) who backs the player establishing the link. The significance of this frequent pattern demonstrates that signed link placement is a complex social process transcending dyads and triads. The fact that the closing player's two reciprocated friends are not connected (highly unlikely in terms of triadic closure dynamics \cite{szell2010multirelational}) suggests that she not only considers her friend, but also the support and status of friends in other social circles.

Figure \ref{fig:pardus3} (right) depicts the second closing pattern. Again this is a balanced closure of a player marking her friend's enemy as an enemy. However, the fourth node represents an already-existing common enemy. This suggests that existing common neighbors in the enemy network influence enemy link placement decisions of friends. This result extends previous insights on the Pardus signed link dynamics which considered only preferential attachment or single triads \cite{szell2010multirelational,szell2010msd}, and it justifies multiplex link prediction algorithms which account for common neighbors in different layers.

\section{Conclusion}
In this paper, we describe a new framework to perform multiplex link prediction via graph association rules. Multiplex link prediction is the task of forecasting new links appearing in a multiplex network, specifying not only which two nodes will connect to each other, but also of which type. We perform a series of experiments showing how this approach outperforms the current state of the art and comes close to an ideal ensemble classifier. We show both quantitative and qualitative improvements by adding new features to multiplex link prediction such as the ability of predicting incoming nodes. In the signed network scenario, we extend classical social balance theory by considering patterns of four nodes, rather than limiting to triangles.

There are a number of future directions to further increase multiplex link prediction performance. First, we can integrate our framework in \texttt{Moss}, combining the mining step with the link prediction step. This will increase time efficiency. We could also perform the experiments on the extended framework with many-to-many interlayer mappings, which we outlined. Finally, we could investigate more scoring schemes rather than relying on the simple rule count weighted by confidence. Despite these possibilities for technical improvements, we proved the usage of graph association rules to be a quantitative and qualitative improvement over previous multiplex link predictors, with unique domain applications.

\bibliographystyle{aaai}
\bibliography{biblio}

\end{document}